\documentclass[useAMS,usenatbib]{mn2e}

\usepackage{epsfig}
\def\kms{$\mbox{km\,s}^{-1}$}

\newcommand{\Oiii}{[{\sc O$\,$iii}]}
\newcommand{\Oii}{[{\sc O$\,$ii}]}
\newcommand{\Ni}{[{\sc N$\,$i}]}
\newcommand{\Hb}{H$\beta$}

\newcommand{\placefigone}{
\begin{figure*}
\includegraphics[width=\textwidth]{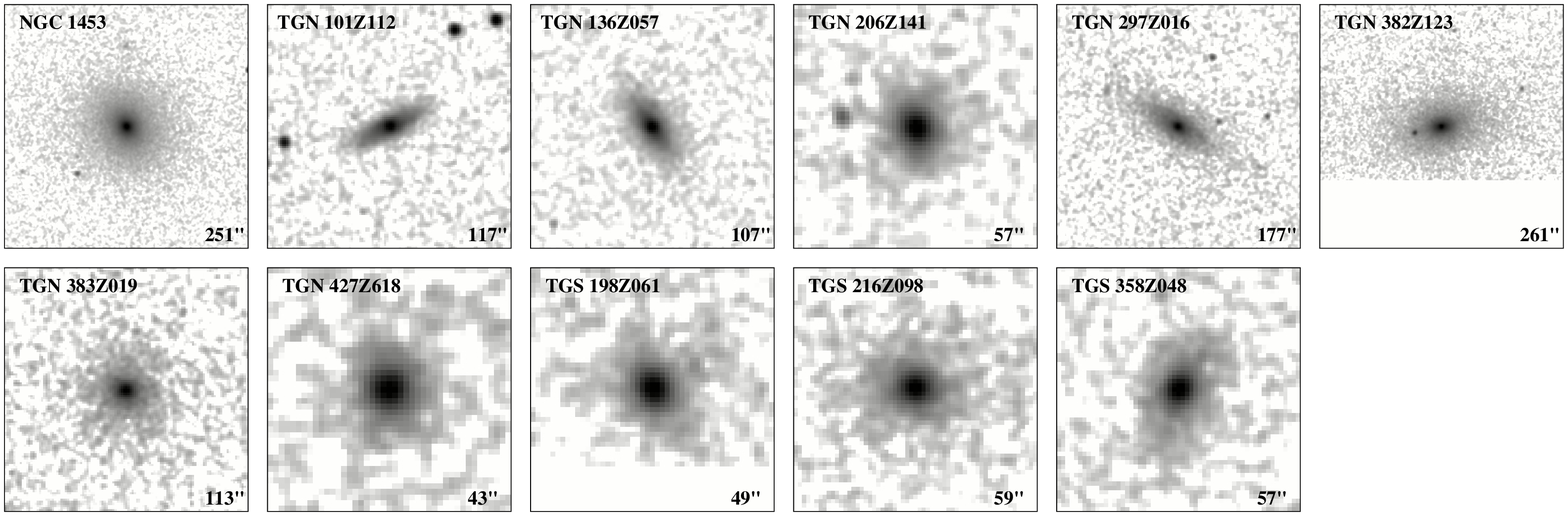}
\caption{2MASS ``postage stamp'' J-band images for the 11 galaxies 
  of the 2dFGRS sample.  In each panel the size of the image is
  indicated in arcsec in the lower right, and north is up and east
  is left.\label{2mass_images}}
\end{figure*}
}

\newcommand{\placefigtwo}{
\begin{figure}
\includegraphics[width=83mm]{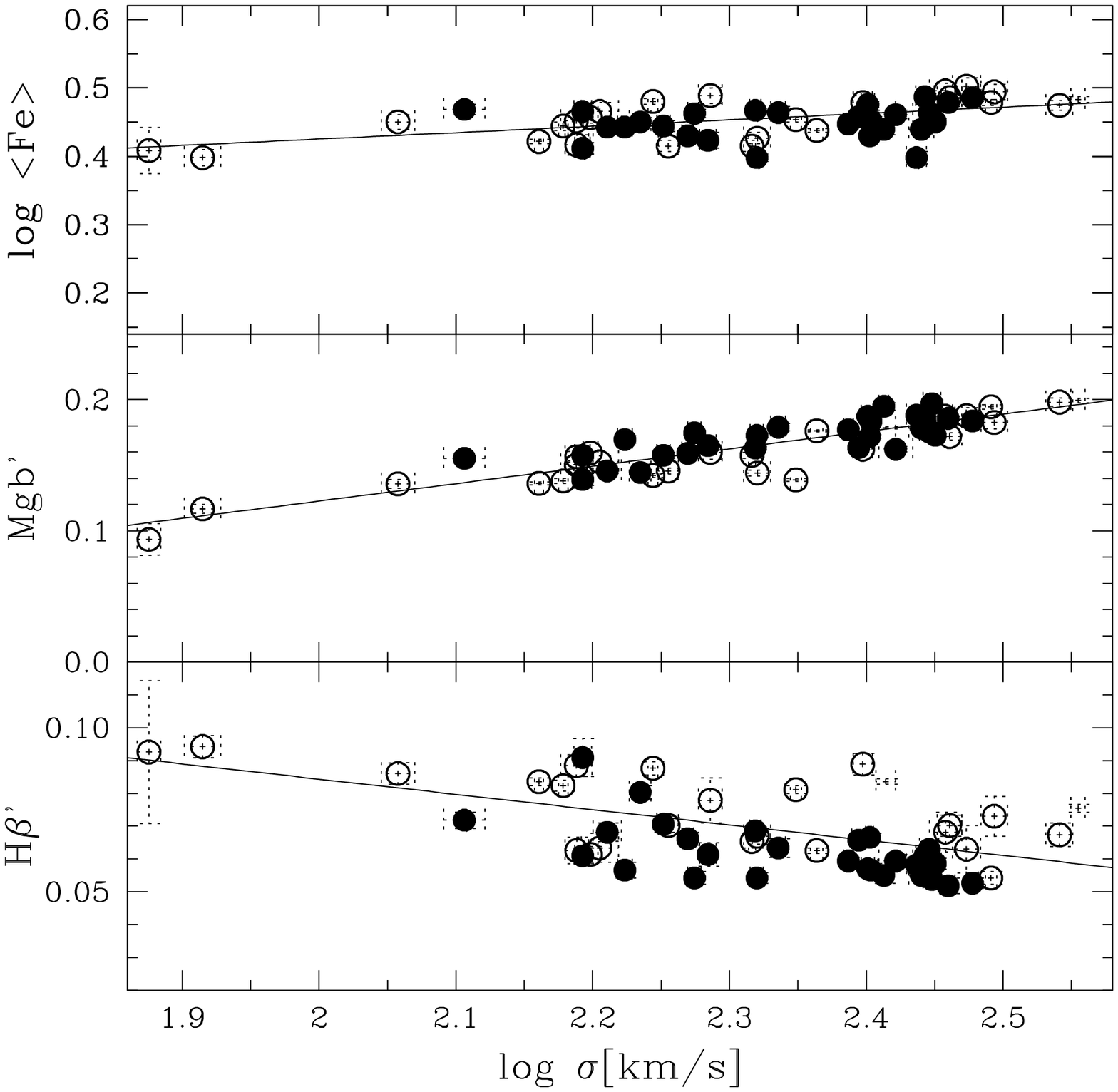}
\caption{ Relation between log ($<$Fe$>$), Mg\,$b\,^\prime$,
  H$\beta\,^\prime$ and $\log \sigma$ for the cluster galaxies from
  \citet{Tho05}. Open circles show the S0s, the filled circles
  represent the elliptical galaxies. The solid lines show a
  least-square fit to the data.\label{indices_sig_cluster}}
\end{figure}
}

\newcommand{\placefigthree}{
\begin{figure*}
\begin{minipage}{85mm}
\includegraphics[width=83mm]{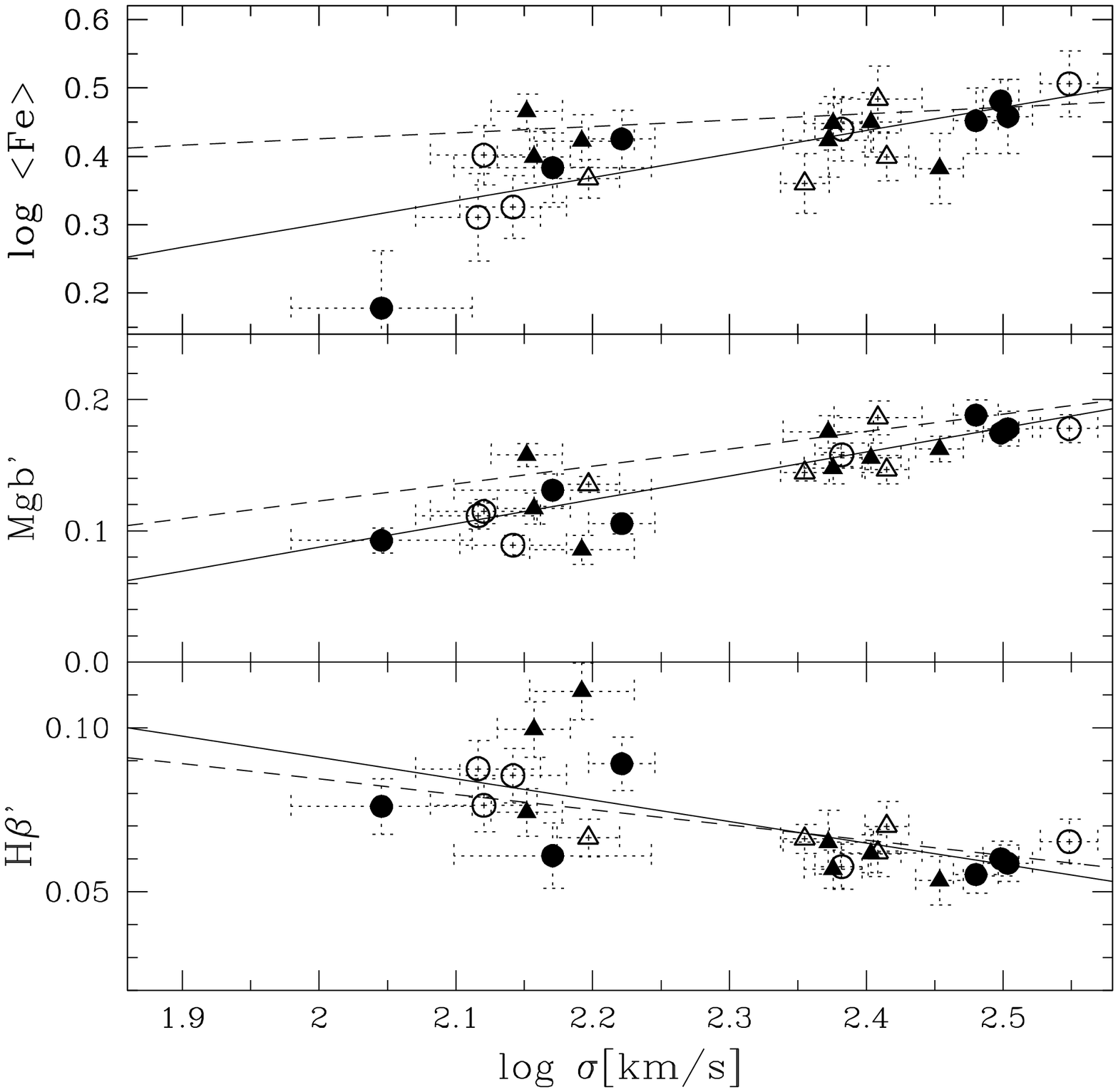}
\end{minipage}
\begin{minipage}{85mm}
\includegraphics[width=83mm]{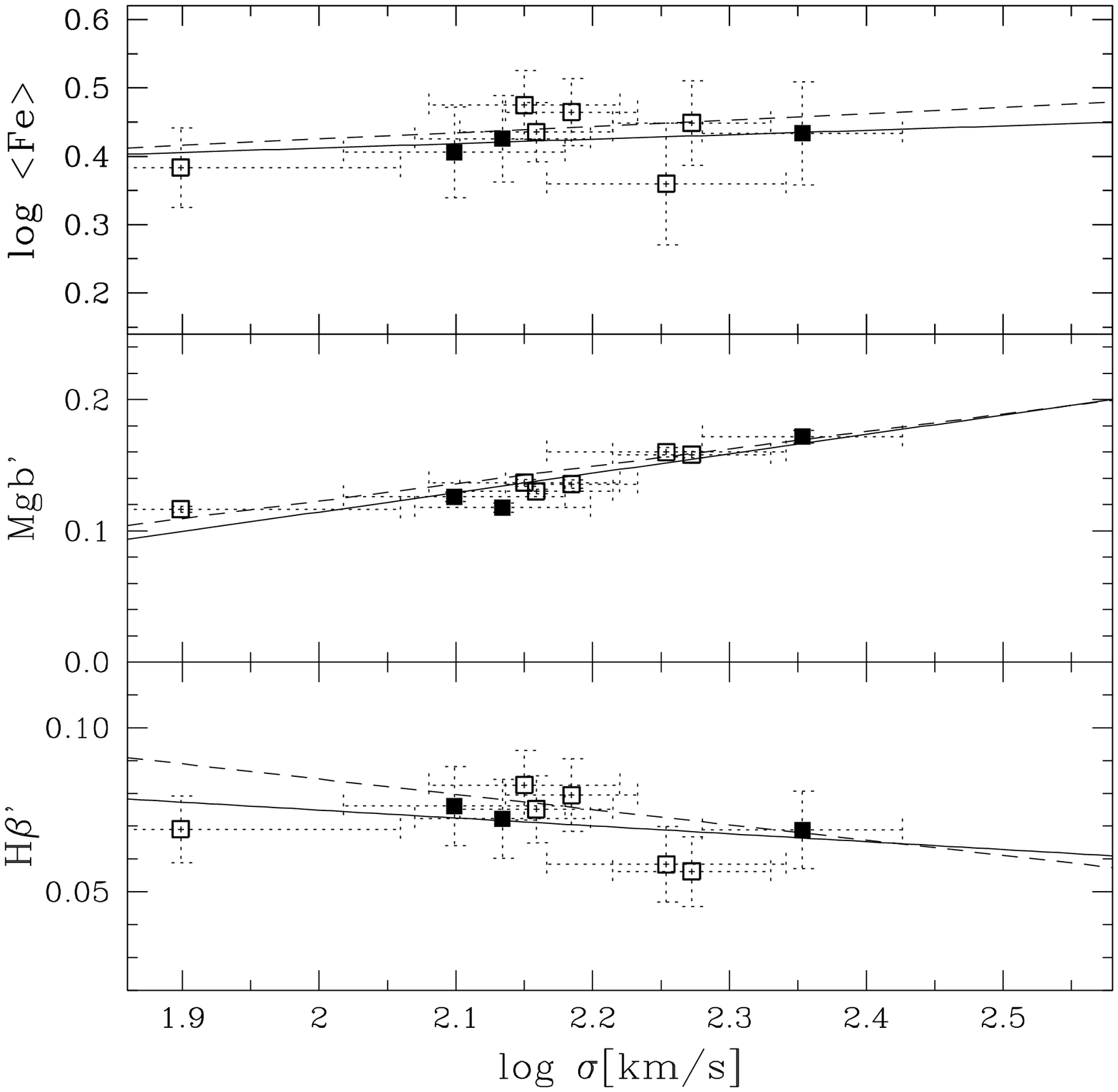}
\end{minipage}
\caption{ Relation between log ($<$Fe$>$), Mg\,$b\,^\prime$,
  H$\beta\,^\prime$ and $\log \sigma$ for the galaxies in low-density
  environments.  In the left plot, the 2dFGRS sample is represented by
  circles, Colbert's sample by triangles. The right plot shows the
  \citet{Kun02} sample in rectangles.  For both plots, the S0 galaxies
  are shown with open symbols, the ellipticals with filled symbols.
  In all panels the linear fit to the the data is shown by the solid
  line, whereas the dashed line shows for comparison the same fit for
  the cluster sample (see Figure~\ref{indices_sig_cluster}).
  \label{indices_sig_ours}}
\end{figure*}
}

\newcommand{\placefigfour}{
\begin{figure}
\includegraphics[width=83mm]{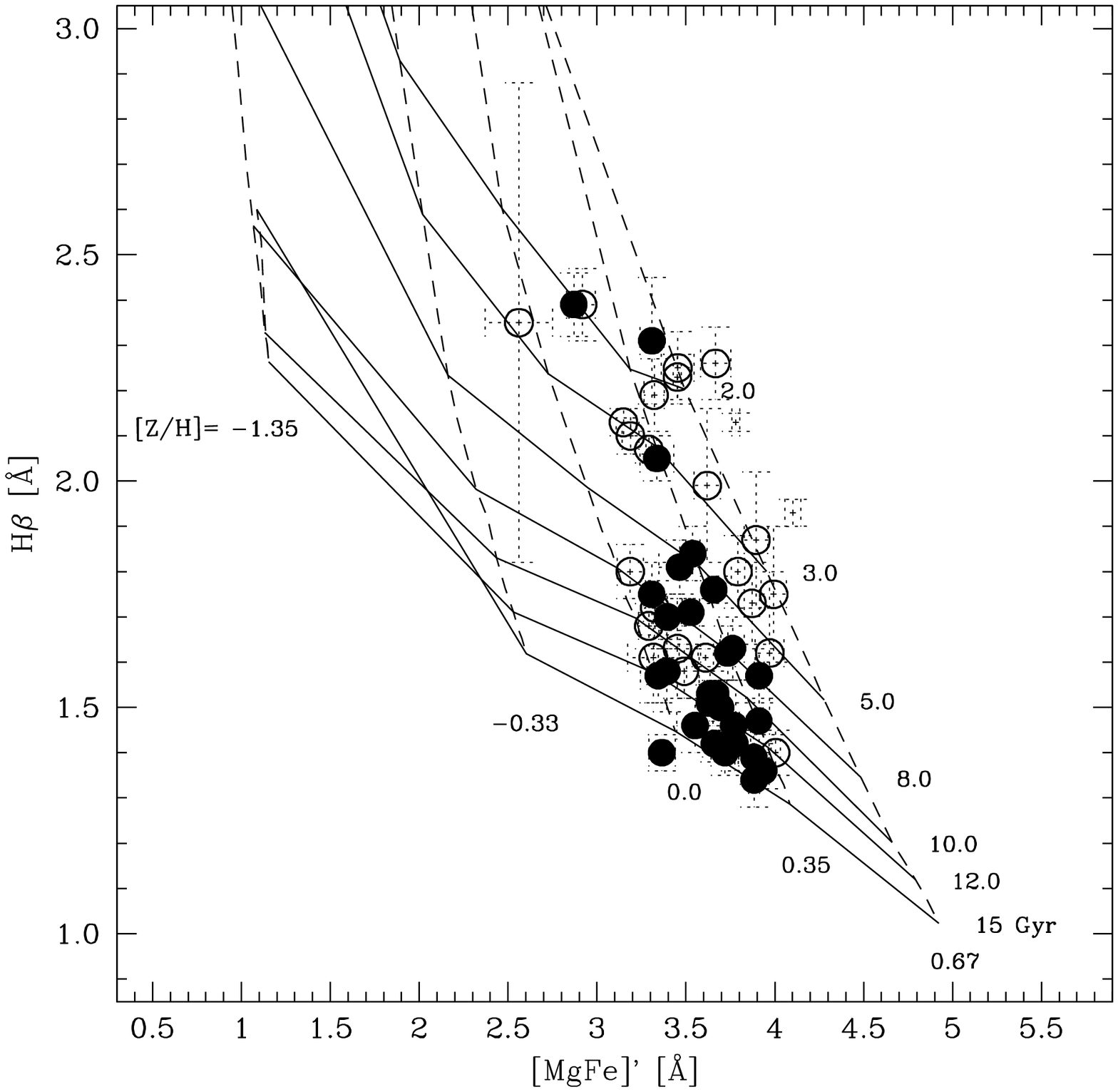}
\caption{The age-sensitive index H$\beta$ as a function of the
  metallicity-sensitive index [MgFe]$^\prime$. Overplotted are
  the stellar population models of \citet{Tho03a}. The solid lines are
  lines of constant age from 2 to 15 Gyr. The dashed, almost vertical,
  lines stand for constant metallicity from [Z/H] $=$ -1.35 to 0.67
  dex. The age and metallicity steps are indicated to the right and at
  the bottom of the model predictions.  The symbol definitions are the
  same as in Figure~\ref{indices_sig_cluster}.
  \label{MgFep_Hb_cluster}}
\end{figure}
}

\newcommand{\placefigfive}{
\begin{figure*}
\begin{minipage}{85mm}
\includegraphics[width=83mm]{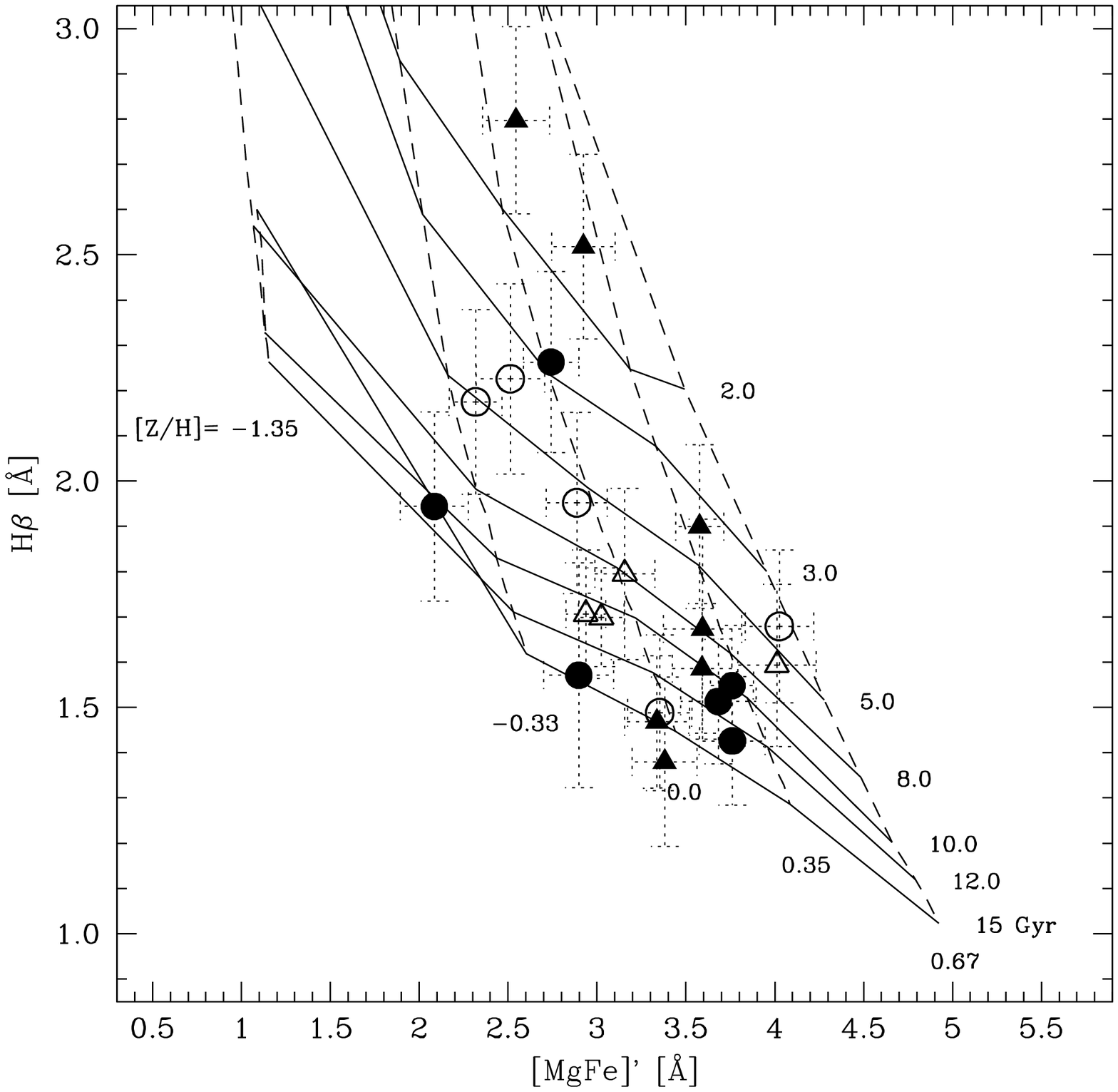}
\end{minipage}
\begin{minipage}{85mm}
\includegraphics[width=83mm]{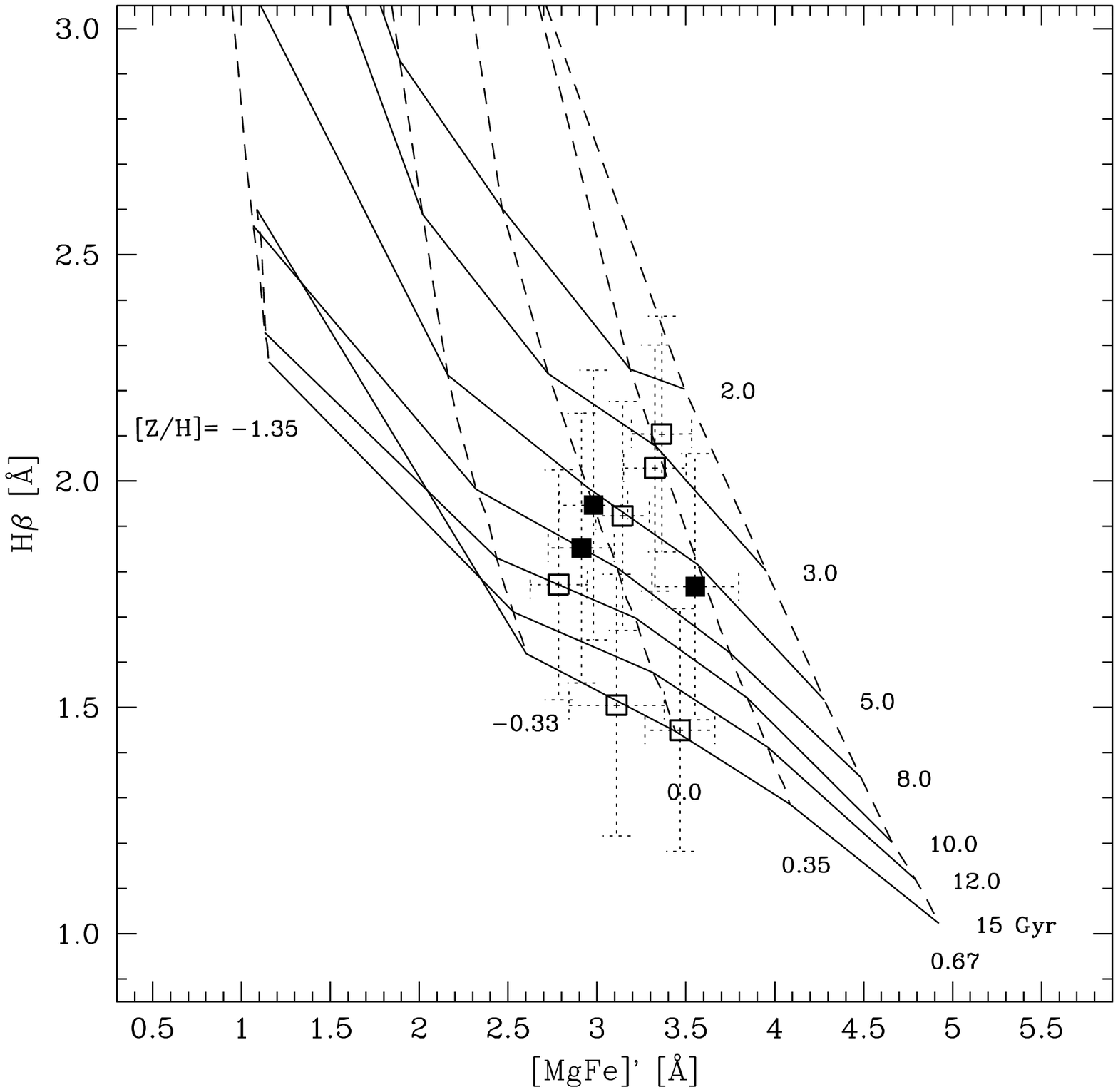}
\end{minipage}
\caption{The Balmer-index H$\beta$ as a function of [MgFe]$^\prime$. 
  The symbols are the same as Figure~\ref{indices_sig_ours}. The left
  plot shows the 2dFGRS and Colbert's sample, the right plot the LDE
  sample \citep{Kun02}. \label{MgFep_Hb_ours}}
\end{figure*} 
}

\newcommand{\placefigsix}{
\begin{figure}
\includegraphics[width=83mm]{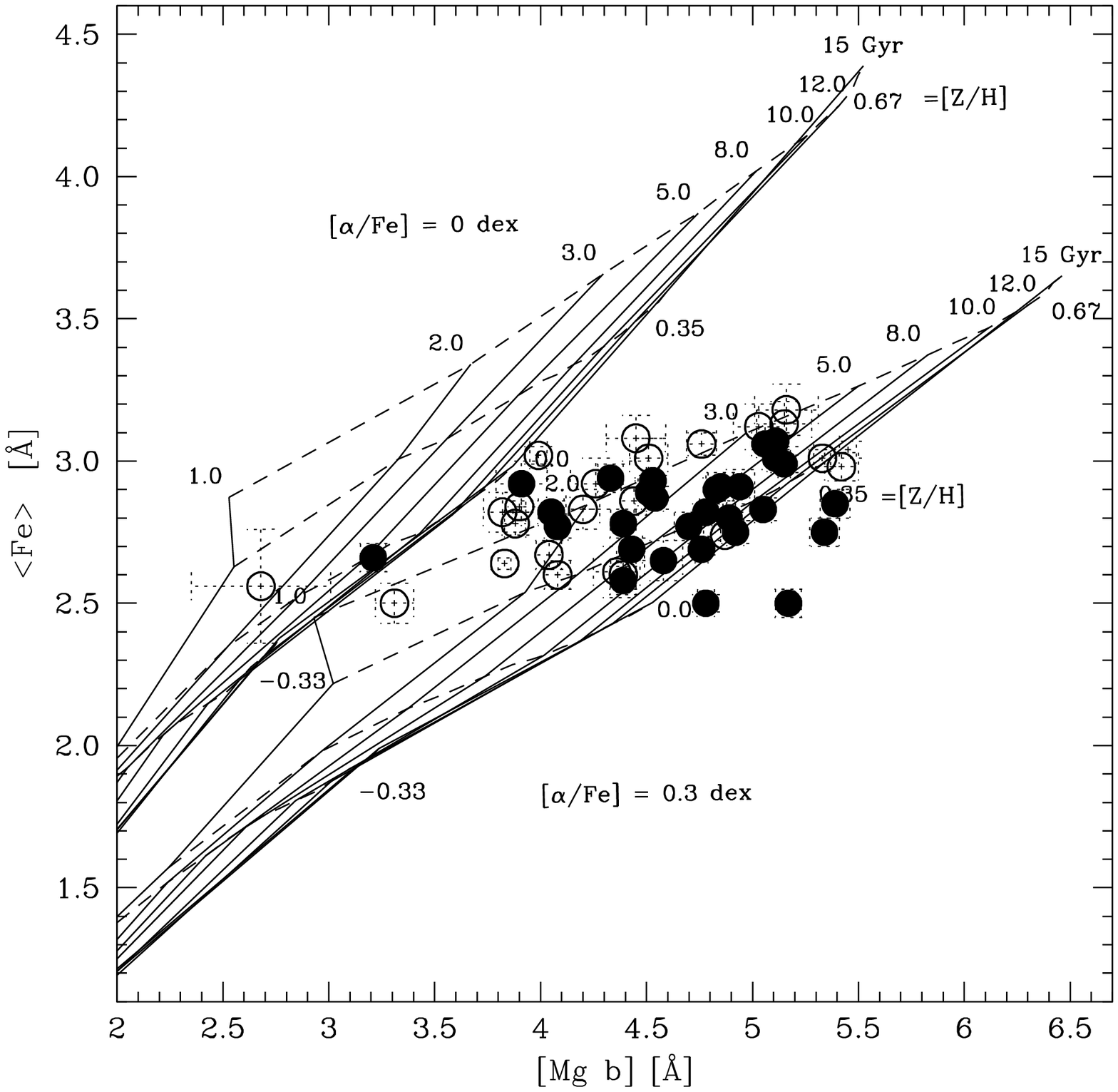}
\caption{Mg\,$b$ versus $<$Fe$>$ for the cluster environments. Open and
  filled circles represent lenticular and elliptical galaxies,
  respectively. Overplotted are SSP models from \citet{Tho03a}
  for solar and non-solar abundance ratios, [$\alpha$/Fe] $=$ 0.0 and
  0.3 dex, respectively.  \label{Mg_Fe_cluster}}
\end{figure}
}

\newcommand{\placefigseven}{
\begin{figure*}
\begin{minipage}{85mm}
\includegraphics[width=83mm]{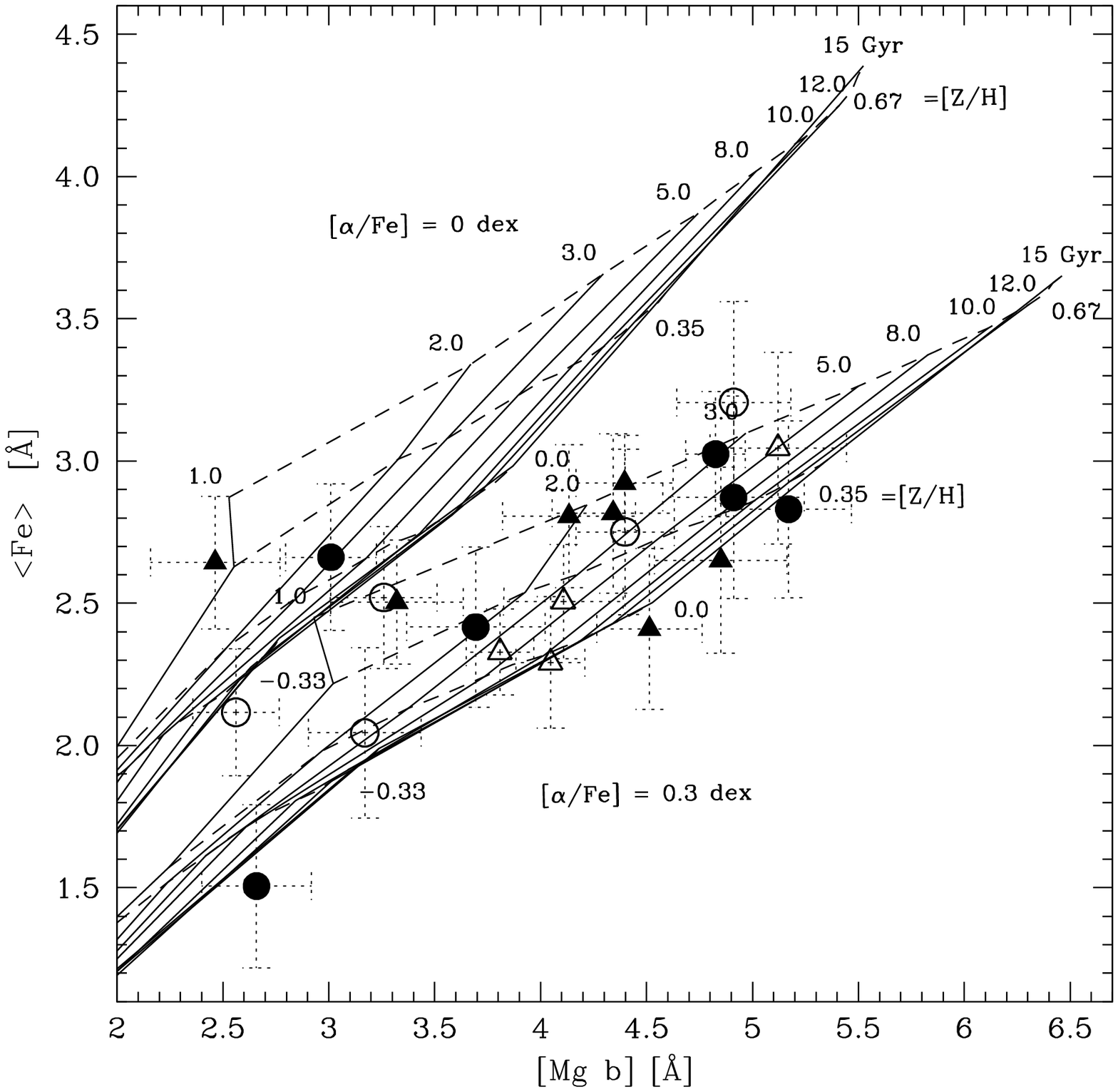}
\end{minipage}
\begin{minipage}{85mm}
\includegraphics[width=83mm]{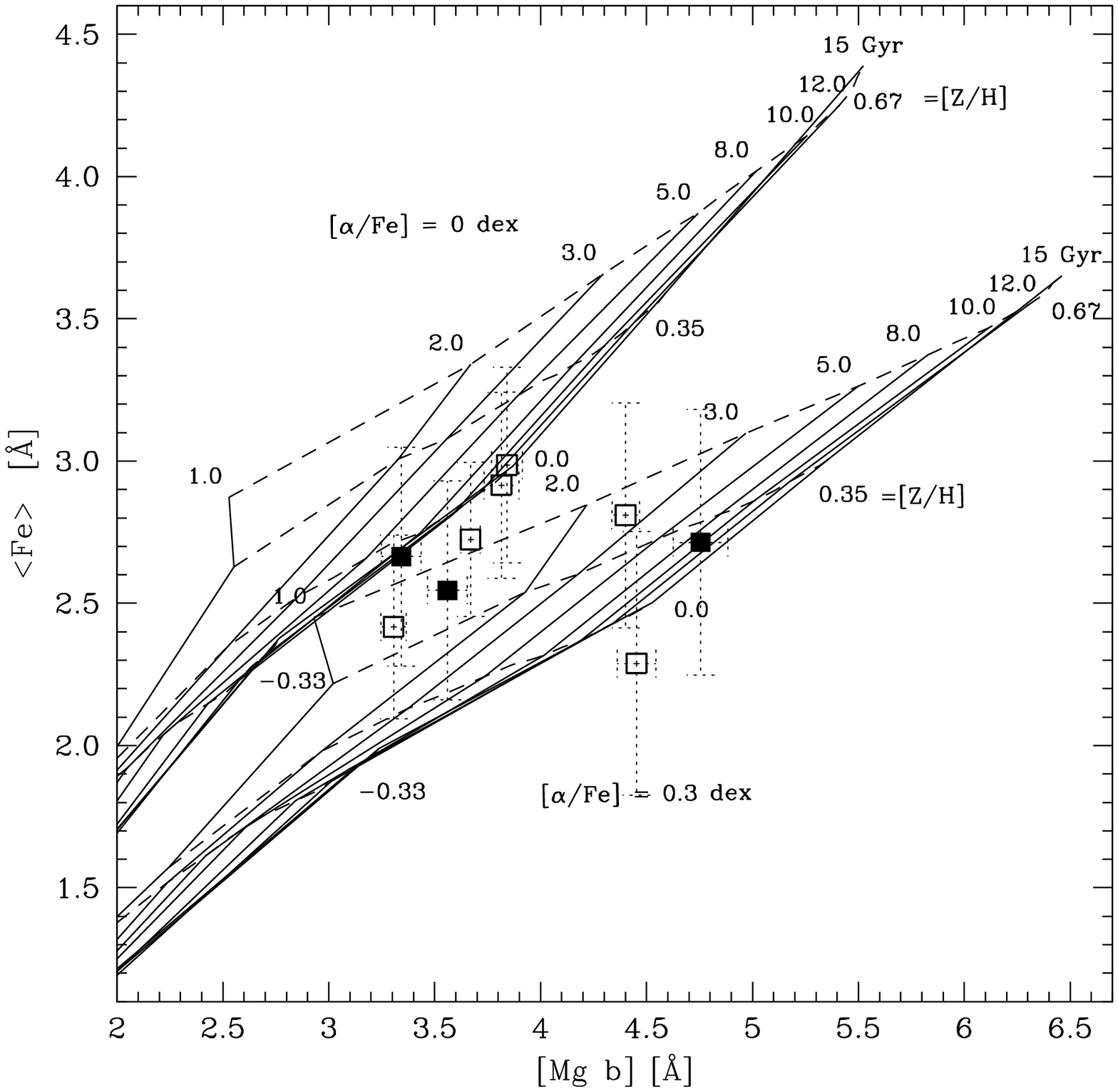}
\end{minipage}
 \caption{The index $<$Fe$>$ as a function of Mg\,$b$, with stellar
   population models overplotted. These are the same than in figure
   \ref{Mg_Fe_cluster}. The symbols are the same than figure
   \ref{indices_sig_ours}.\label{Mg_Fe_ours}}
\end{figure*}
}

\newcommand{\placefigeight}{
\begin{figure*}
\begin{minipage}{85mm}
\includegraphics[width=83mm]{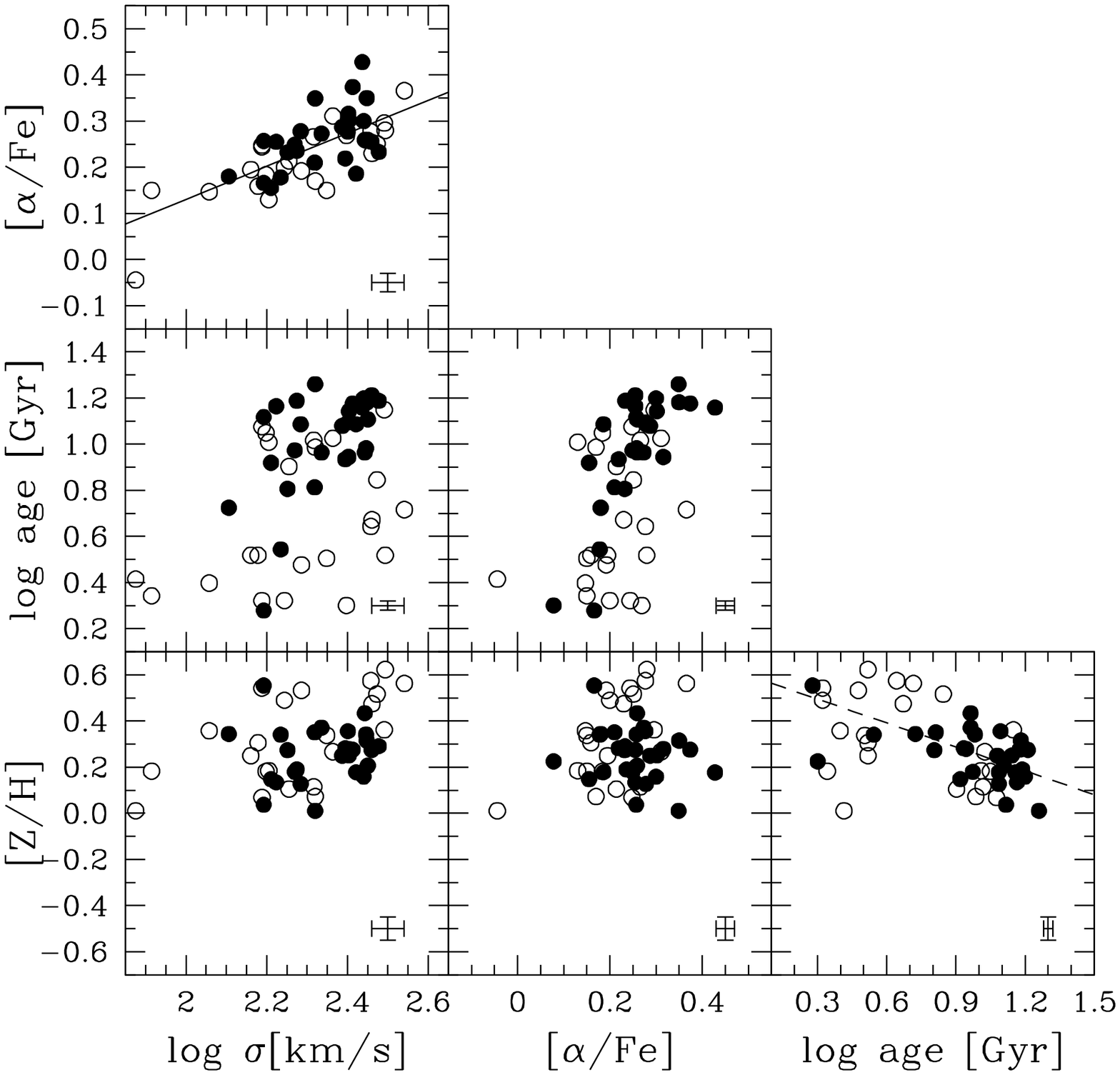}
\end{minipage}
\begin{minipage}{85mm}
\includegraphics[width=83mm]{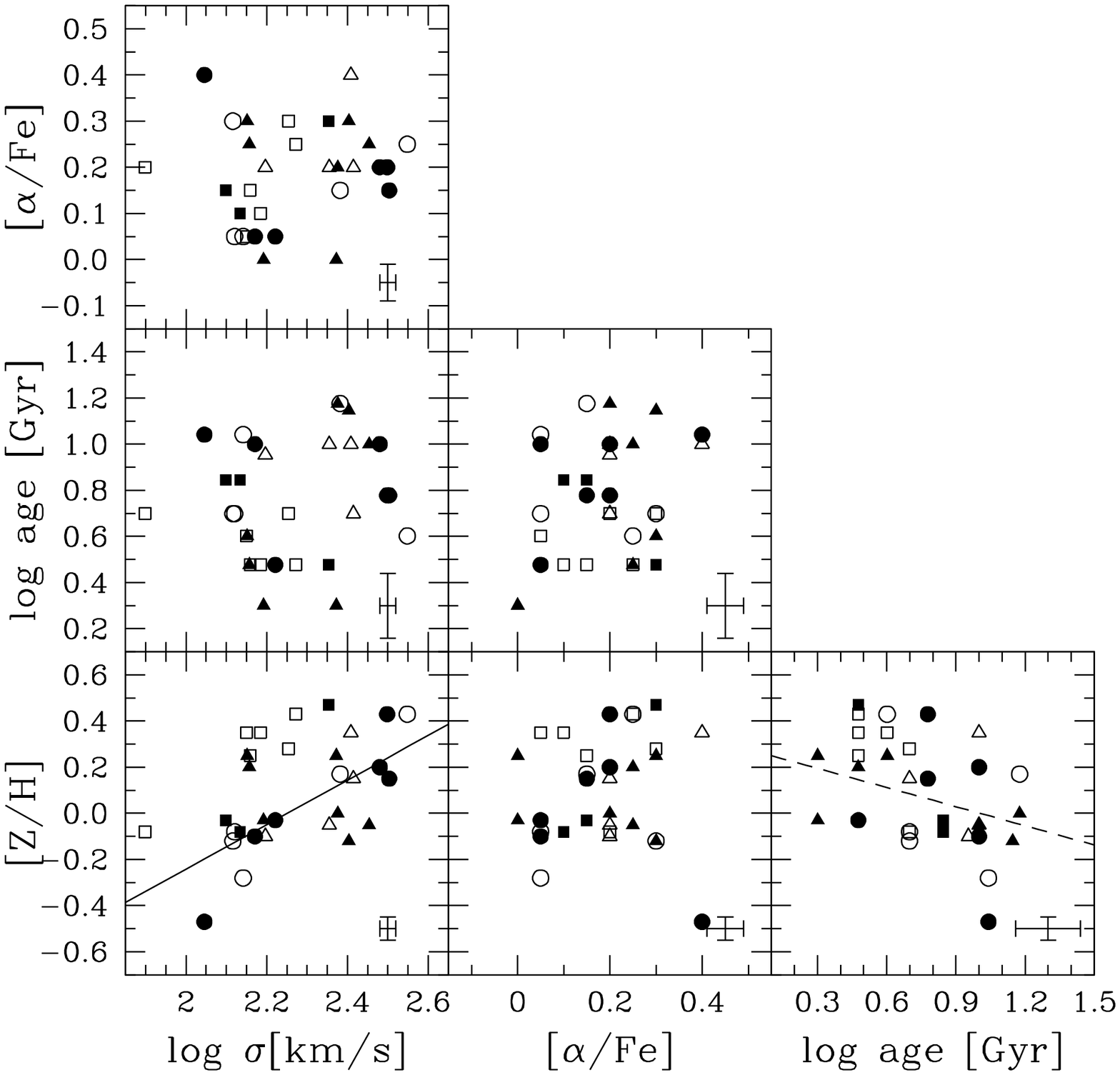}
\end{minipage}
 \caption{ The global relations between log age, [Z/H], [$\alpha$/Fe]
   and log $\sigma$ are shown. The symbols are the same than figure
   \ref{indices_sig_ours}, and figure \ref{indices_sig_cluster}. The
   left plot shows the clusters galaxies, the right panel combines the
   2dFGRS and Colbert's samples with the LDE sample. The solid
   lines identify a least-square fit between the parameters. Typical
   error bars are shown at the lower-right corner of each diagram.
   These do not account for the fact that the errors on age, [Z/H] and
   [$\alpha$/Fe] can be correlated, which can impact the significance
   of the observed correlations between these parameters. In
   particular, this can be the case for the age-[Z/H] relations, which
   are shown by dashed lines.
\label{proctor_plots_cluster_ours}} 
\end{figure*}
}

\newcommand{\placetabone}{
\begin{table*}
 \centering
\begin{minipage}{145mm}
  \caption{Observed galaxies in low-density environments: 2dFGRS and \citet{Colb01} sample, and calibration Lick galaxies. \label{table1}}
  \begin{tabular}{lrrccccc}\\
 \hline
Name        & RA (B1950)                   & DEC (B1950)                  &b$_J$ & Type  & $cz$ & $\sigma$  & Run \\
(1) & (2) & (3) & (4) & (5) & (6) & (7) & (8) \\
\hline
\multicolumn{8}{c}{2dFGRS sample} \\
\hline
NGC\,1453     &  3$^{h}$ 43$^{m}$ 57\farcs4 &  -4\degr 07$^{m}$ 24\arcsec & 12.77 & E2-3 & 3934$\pm$ 9 & 314 $\pm$ 10 & dec 2000 \\
TGN\,101Z112  & 10$^{h}$ 36$^{m}$ 36\farcs5 &  -5\degr 14$^{m}$ 29\arcsec & 15.03 & S0   &  8330 $\pm$ 11  & 241 $\pm$ 11&  apr 2001\\
TGN\,136Z057  & 13$^{h}$ 22$^{m}$ 18\farcs0 &  -3\degr 49$^{m}$ 24\arcsec & 14.44 & S0   & 6132  $\pm$  18 & 353 $\pm$  17 & apr 2001 \\
TGN\,206Z141  & 14$^{h}$ 08$^{m}$ 06\farcs0 &  -2\degr 26$^{m}$ 57\arcsec & 15.50  & E   & 8093  $\pm$  8 & 166 $\pm$  9 & apr 2001  \\
TGN\,297Z016  & 10$^{h}$ 49$^{m}$ 33\farcs6 &  -0\degr 17$^{m}$ 47\arcsec & 14.51 & S0/a & 5515 $\pm$  9 & 138 $\pm$  12 &  apr 2001\\
TGN\,382Z123  & 12$^{h}$ 01$^{m}$ 54\farcs0 &  +2\degr 10$^{m}$ 22\arcsec & 14.36 & E    & 5871  $\pm$  11 & 302 $\pm$  11 &  apr 2001\\
TGN\,383Z019  & 12$^{h}$ 08$^{m}$ 31\farcs0 &  +1\degr 15$^{m}$ 06\arcsec & 14.21 & S0-  &6092 $\pm$  10 & 130 $\pm$  14 & apr 2001  \\
TGN\,427Z618  & 10$^{h}$ 32$^{m}$ 16\farcs0 &  +0\degr 56$^{m}$ 06\arcsec & 15.71  & E   & 9502 $\pm$  11 & 111 $\pm$  17 & apr 2001\\
TGS\,198Z061  & 23$^{h}$ 59$^{m}$ 01\farcs5 & -27\degr 51$^{m}$ 08\arcsec & 15.36 & E      & 8021 $\pm$  22 & 148 $\pm$  24 & dec 2000 \\
TGS\,216Z098  &  1$^{h}$ 22$^{m}$ 49\farcs6 & -27\degr 30$^{m}$ 24\arcsec & 15.14 & E & 9330  $\pm$  10 &  214 $\pm$   13  & dec 2000 \\
TGS\,358Z048  &  0$^{h}$ 02$^{m}$ 58\farcs2 & -30\degr 51$^{m}$ 55\arcsec & 15.37 & S0     & 8330  $\pm$  11 & 132 $\pm$  11 & dec 2000\\
\hline
\multicolumn{8}{c}{Galaxies from Colbert et al. (2001)} \\
\hline
A0718-34    &  7$^{h}$ 18$^{m}$ 56\farcs2 & -34\degr 01$^{m}$ 25\arcsec & 15.8  & SA0-  & 8500  $\pm$  15 & 256  $\pm$  19 & dec 2000\\ 
ESO\,065-G001& 12$^{h}$ 34$^{m}$ 05\farcs0 & -72\degr 18$^{m}$53\arcsec  & 14.50 & E4 & 7099 $\pm$  15 & 236 $\pm$  18  & apr 2001\\
ESO\,505-G015& 12$^{h}$ 04$^{m}$ 33\farcs0 & -25\degr 24$^{m}$ 53\arcsec & 14.55 &E+3 & 7451 $\pm$  10 & 284  $\pm$  11 & apr 2001\\
ESO\,574-G017& 12$^{h}$ 37$^{m}$ 58\farcs0 & -20\degr 17$^{m}$ 15\arcsec & 15.22 &E+ & 8428  $\pm$  10 & 156 $\pm$  13 & apr 2001\\
IC\,1156     & 15$^{h}$ 58$^{m}$ 25\farcs0 & +19\degr 51$^{m}$ 44\arcsec & 14.48 & E  & 9213 $\pm$  9 & 234 $\pm$  11 & apr 2001\\ 
IC\,2980     & 11$^{h}$ 54$^{m}$ 59\farcs9 & -73\degr 24$^{m}$ 24\arcsec & 14.6  & E3  & 2092 $\pm$  6 & 141 $\pm$  8 & dec 2000 \\ %8352 !!
IIZW017     &  4$^{h}$ 32$^{m}$ 42\farcs0 &  -1\degr 50$^{m}$ 00\arcsec & 15.6  & cE pec &  9728 $\pm$  8 &143 $\pm$  9  & dec 2000\\  
NGC\,179     &  0$^{h}$ 35$^{m}$ 15\farcs9 & -18\degr 07$^{m}$ 27\arcsec & 14.30 & SAB0- & 5976 $\pm$ 9 & 260 $\pm$ 9 & dec 2000\\
NGC\,1132     &  2$^{h}$ 50$^{m}$ 19\farcs6 &  -1\degr 28$^{m}$ 45\arcsec & 13.25 & E  & 6935  $\pm$  11 & 253 $\pm$  13 & dec 2000 \\
NGC\,3332     & 10$^{h}$ 37$^{m}$ 51\farcs0 &  +9\degr 26$^{m}$ 40\arcsec & 13.34 & (R)SA0- & 5758  $\pm$ 7 & 226 $\pm$ 9 & dec 2000 \\
NGC\,6799    & 19$^{h}$ 28$^{m}$ 11\farcs0 & -56\degr 00$^{m}$ 53\arcsec & 13.39 & SA0- & 3363  $\pm$  6 & 157 $\pm$  8 &  apr 2001\\
\hline
\multicolumn{8}{c}{Lick galaxies} \\
\hline
NGC\,1600     &  4$^{h}$ 29$^{m}$ 11\farcs7 &  -5\degr 11$^{m}$ 38\arcsec & 11.93 & E3 & 4681  $\pm$ 8 & 315 $\pm$ 8  & dec 2000 \\
NGC\,1700     &  4$^{h}$ 54$^{m}$ 28\farcs6 &  -4\degr 56$^{m}$ 33\arcsec & 12.20 & E4 & 3889  $\pm$ 7  & 246 $\pm$  7 & dec 2000 \\  
\hline \\ 
\end{tabular}

\medskip

Notes: The first column lists the name of the galaxy, columns (2) and
(3) list the B1950 coordinates, column (4) shows the (total) b$_J$
magnitude, column (5) the type of the galaxy as found in NED or the
SDSS. Columns (6) and (7) list the recession velocity and the velocity
dispersion, respectively, as measured from our spectra. Column (8)
indicates in which run the galaxy was observed.
\end{minipage}

\end{table*}
}

\newcommand{\placetabtwo}{
\begin{table}
\caption{The instrumental set-up. \label{table2}}
\begin{tabular}{cc}\\
 \hline
Telescope      & MSSSO (2.3m)\\
Dates          & 16--20~Dec 2000, 26~Apr -- 1~May 2001\\
Instrument     & DBS spectrograph (blue arm)\\
 \hline
Spectral range & 3690 \AA--5600\AA\\
Grating        & 600 lines mm$^{-1}$\\
Dispersion     & 1.1 \AA$\,$pixel$^{-1}$\\
Resolution     & $\approx$ 2.3 \AA\\
Spatial Scale  & 0$\farcs$91 pixel$^{-1}$\\
Slit Width     & 2$\farcs$0\\
Detector       & SITe (1752 $\times$ 532 pixels; 15 $\times$ 15 $\micron$m)\\
Gain           & 1.0 e$-$ ADU$^{-1}$\\
Read-out-Noise & 5.5 e$-$ (rms)\\
Typical seeing & 1$\farcs$5\\
\hline
\end{tabular}
\end{table}
}

\newcommand{\placetabthree}{
\begin{table*}
\centering
\begin{minipage}{170mm}
\caption{Lick/IDS indices and derived stellar population parameters for
  the low-density environment galaxies \label{index-1}}
\hspace*{-0.4cm}
\begin{tabular}{lrrrrrrrrrrrrr}
\hline\hline
Galaxy       & age & [Z/H] & [$\alpha$/Fe] & CN$_1$ & CN$_2$ & Ca4227 & G4300 & Fe4383 & H$\beta$ & Fe5015 & Mg$b$ & Fe5270 & Fe5335 \\ 
             & Gyr &       &               & [mag]  & [mag]  & [\AA]  & [\AA] & [\AA]  & [\AA]    & [\AA]  & [\AA] & [\AA]  & [\AA]  \\ 
\hline
  NGC\,1453 &   6.0 &  0.43 & 0.20 &  0.116 & 0.156&  0.92 &  5.96 &  5.90 &  1.55&  5.33 &  4.83 &  3.24 &  2.81 \\
      $\pm$ &   2.8 &  0.10 & 0.04 &  0.003 & 0.004&  0.14 &  0.11 &  0.31 &  0.13&  0.76 &  0.14 &  0.17 &  0.41 \\
 TGN\,101Z112 &  15.0 &  0.17 & 0.15 &  0.094 & 0.136&  1.18 &  6.66 &  6.34 &  1.49&  5.41 &  4.40 &  3.20 &  2.31 \\
      $\pm$ &   1.9 &  0.08 & 0.06 &  0.006 & 0.007&  0.17 &  0.20 &  0.43 &  0.17&  0.86 &  0.23 &  0.27 &  0.52 \\
 TGN\,136Z057 &   4.0 &  0.43 & 0.25 &  0.122 & 0.163&  2.25 &  5.77 &  4.93 &  1.68&  5.34 &  4.91 &  2.99 &  3.42 \\
      $\pm$ &   2.4 &  0.13 & 0.05 &  0.004 & 0.005&  0.20 &  0.16 &  0.41 &  0.17&  1.01 &  0.27 &  0.25 &  0.66 \\
\hline
\end{tabular}

\medskip

Notes: The full table is available in the electronic version of MNRAS.

\end{minipage}

\end{table*}

}

\title[Central stellar populations of early-type galaxies in
low-density environments]{Central stellar populations of early-type
  galaxies in low-density environments}

\author[Maela Collobert et al.]{Maela Collobert $^{1}$\thanks{E-mail :
    maelasc@astro.ox.ac.uk}, Marc Sarzi$^{1,2}$, Roger L. Davies$^{1}$,
  Harald Kuntschner$^{3}$,
  \newauthor Matthew Colless$^{4}$ \\
  $^{1}$University of Oxford, Astrophysics, Denys Wilkinson Building,
  Keble Road, Oxford, OX1 3RH, UK\\
  $^{2}$Centre for Astrophysics Research, University of Hertfordshire, College Lane, Hatfield, Herts, AL10 9AB, UK\\  
  $^{3}$Space Telescope European Coordinating Facility, ESO,
  Karl-Schwarzschild-Str. 2, 85748 Garching, Germany \\
  $^{4}$Anglo-Australian Observatory, PO Box 296, Epping, NSW 2111,
  Australia}

\begin{document}
\pagerange{\pageref{firstpage}--\pageref{lastpage}} \pubyear{2004}
\maketitle

\label{firstpage}

\begin{abstract}

Following the pilot study of \citet{Kun02} we have investigated the
properties of a volume and magnitude limited ($cz<10,000\,$\kms, b$_J
<16$) sample of early type galaxies that were carefully selected from
the AAO two degree field galaxy redshift survey to have no more than
one and five companions within 1 and 2\,Mpc, respectively. We used
images from the DSS to confirm the E/S0 morphologies.
We augmented this sample with field galaxies from \citet{Colb01}
selected as having no neighbour within 1\,Mpc and $\pm
1\,000$\,\kms. We present spectroscopic observations of 22 galaxies
from the combined sample, from which central velocity dispersions and
the Lick stellar population indices were measured.  After carefully
correcting the spectra for nebular emission we derived
luminosity-weighted ages, metallicities, and $\alpha$-element
abundance ratios. We compare these isolated galaxies with samples of
early-type galaxies in the Virgo and Coma clusters, and also with the
previous sample of galaxies in low-density regions of \citet{Kun02}.
We find that galaxies in low-density environments are younger and have
a greater spread of ages compared to cluster galaxies. They also show
a wider range of metallicities at a given velocity dispersion than
cluster galaxies, which display only super-solar metallicities.
On average cluster, as well as, isolated galaxies show non-solar
abundance ratios in $\alpha$-elements, suggesting that, independent of
galactic environment, star formation occurred on short
time-scales. However, the abundance ratios for our low-density
environment sample galaxies do not scale with the stellar velocity
dispersion as observed in clusters. In fact we detect a large spread
at a given velocity dispersion even reaching solar abundance ratios.
The metallicity of isolated early-type galaxies is found to correlate
weakly with $\sigma$.
We reason that early-type galaxies in low-density environments
experienced merging-induced star-formation episodes over a longer and
more recent period of time compared to a cluster environment, and
speculate that a considerable fraction of their stars formed out of
low-metallicity halo gaseous material during the slow growth of a
stellar disk between merging events.

\end{abstract}
\placetabone


\section{Introduction}

Understanding the formation and evolution of early-type galaxies is
one of the major challenges for current structure-formation models. In
particular, one key issue to be addressed is whether the environmental
impact on the formation and evolution of galaxies predicted by N-body
and semi-analytical models \citep[e.g.,][]{Bau96,Kau99,Spr01,Kho03} is
observed in real galaxies.
In the framework of a cold-dark matter (CM) universe, structure
assembles hierarchically by successive mergers of smaller
structures. Observations of the galaxy merging rate in different
environments suggest that interactions in high-density environments
were more frequent than interactions in the field at high
redshifts. Conversely interaction rates in the field are higher at low
redshifts \citep{vDo99,lFe00}. These trends are predicted by the CDM
paradigm \citep[e.g.,][]{Gov99,Got01,Kho01}.
For instance, in high-density environments rich clusters form quickly
and subsequent merging between galaxies is suppressed by the large
relative velocities. On the other hand, in low-density environments
galaxies are still assembling today.
The details of these predictions depend on a number of parameters,
such as AGN or supernovae feedback processes, gas cooling and
accretion rates, and, particularly in clusters, tidal interactions and
ram-pressure gas stripping.

In order to constrain the models it is crucial to study galaxies in
{\it both}\/ high- and low-density environments. In particular, we need
to measure the age, metallicity, and $\alpha$-element abundance
  ratio of their main stellar constituents.
However, the stellar population properties of elliptical and
lenticular galaxies in low-density environments have received little
attention because of the difficulty in identifying a reasonable
sample.
Based on a sample of 9 galaxies with less than two neighbors within
$\sim$1.3\,Mpc, \citet[][hereafter K02]{Kun02} found that galaxies in
low-density environments are on average younger, have higher
metallicity, and exhibit gas emission more frequently than similar
galaxies in the Fornax cluster. The presence of young stellar
populations could explain why isolated galaxies seem to fall below
the fundamental plane and have lower mass-to-light ratios
\citep{Red05}.

Following the pilot study of K02, we set out to identify a larger
sample of early-type galaxies in low-density regions and to study
their stellar populations.
We compute the Lick-indices and look for correlations with the
velocity dispersion. We correct the absorption indices, in particular
the Balmer indices, for the biases introduced by the presence of
emission, using the emission-line removal method of \citet{Sar05}. We
use the single stellar population (SSP) models of \citet[][hereafter
TMB03]{Tho03a}, to derive ages, metallicities and
$\alpha$-abundance ratios for our sample galaxies.

\placefigone 
This paper is organized as follows. In section 2 we describe the
selection of our sample. In section 3 we present the observations and
the basic data reduction, along with the methods used to measure the
galaxy recession velocity, central velocity dispersion, emission-line
fluxes, and Lick absorption-line indices. In section 4, we compare
different index combinations with SSP models to derive the
luminosity-weighted mean stellar age, metallicity, and abundance ratio
of our sample galaxies. Finally in section 5 we discuss our results and
draw our conclusions.

\section{Sample selection}

Our sample consists of two sub-samples. The first include galaxies in
low-density environment that were selected using the AAO two degree
field galaxy redshift survey \citep[2dFGRS,][]{Coll01} and the
other is drawn from the sample of isolated galaxies of \citet{Colb01}.
The galaxies from the 2dFGRS survey were carefully chosen using well
defined criteria. We first selected galaxies with recession velocity
$cz<10,000\,$\kms\ and apparent magnitude b$_J<16$, corresponding to a
limiting absolute magnitude of $M_{B}\le-19.13$. This resulted in a
volume and magnitude limited sample of 443 candidate galaxies. For
each candidate, we then determined the number of companion galaxies
within a sphere of radius 1 and 2\,Mpc h$^{-1}$ at least as
bright as the candidate. When no velocity was available for a
companion, we assumed it was at the same distance as the candidate. In
this way we are conservatively overestimating the number of nearby
companions.  We selected only candidate galaxies with no more than one
and five companions within 1 and 2\,Mpc$^{-1}$, respectively, leaving
around 200 candidates.
We further restricted the number of candidates by keeping only objects
with 2dFGRS spectra of sufficient quality to confidently exclude the
presence of strong emission lines. We checked the morphology of these
candidates with the Digital Sky Survey (DSS) and NASA/IPAC
Extragalactic Database (NED) and eliminated those with obvious spiral
or disturbed morphologies, ending up with about thirty galaxies.
Of these, we observed 11 galaxies from the Siding Spring Observatory in
Australia (see also Section~\ref{sec:obs}). 
The morphology of the 2dFGRS targets was further checked a posteriori
using images from the 2MASS survey, which are shown in
Figure~\ref{2mass_images}.

We added to this sample of galaxies in low-density environments 11
isolated galaxies taken from the sample of \citet{Colb01}. They were
selected as isolated objects from the Third reference Catalogue of
Bright Galaxies (RC3) with no cataloged galaxies with known redshift
within a projected radius of 1\,Mpc h$^{-1}$ and a velocity of $\pm\,$
1000\,\kms.  Furthermore, a maximum recession velocity of $cz =
10\,000$\,\kms\ was imposed.

Table~\ref{table1} lists the observed sample galaxies, along with their
main characteristics. We note that the Hubble classification for the
galaxies in our 2dFGRS sub-sample should be considered with care, as it
is based on photographic plates which makes it difficult to
differentiate between lenticular and elliptical galaxies.
Although the isolation criterion applied to select our 2dFGRS galaxies
is less stringent than the one used by \citet{Colb01}, for simplicity
we refer hereafter to our sample galaxies as isolated, to distinguish
it from the low-density environment sample of K02.

\placetabtwo
\section{Observation and data reduction}
\subsection{Observational techniques}
\label{sec:obs}

The spectra were obtained at the ANU 2.3-m telescope at the Siding
Spring Observatory, Australia in two runs: December 2000 and April
2001. The first run was entirely clear whereas in the second run we got
only 3 nights of data. In both runs we used the Double Beam
Spectrograph \citep{Rod88}, the same instrument as K02. Our analysis is
based on data from the blue arm. The spectra obtained with the red
  arm, which covers the 8100--9050\,\AA\/ wavelength region, were not
  used. The details of the set-up are given in Table~\ref{table2}.

The seeing was in the range $\approx$ 1$\farcs$3 -- 2$\farcs$0. A
typical exposure time for galaxies was 3600s, subdivided in multiple
exposures bracketed by neon-argon lamp calibration spectra, for
wavelength calibration.
Spectrophotometric standard stars were also observed to calibrate the
response function of the system. For comparison purposes we observed
two galaxies from the Lick sample, they are listed in
Table~\ref{table1}. For the brighter galaxies, such as those from the
Lick samples just one exposure was taken. The resulting rest-wavelength
range is 3690--5600\,\AA, covering a range of Balmer and metallic
absorption lines such as H$\beta$, H$\gamma$, Mg\,$b$, Fe5270, Fe5335,
and the \Oii$\lambda3727$ and \Oiii$\lambda\lambda4959,5007$ emission
lines.

\subsection{Basic data reduction}

The data reduction used standard IRAF software packages. Each frame was
overscan-corrected and bias subtracted. Both skyflats and domeflats
were used to correct for the pixel-to-pixel sensitivity variations and
illumination correction along the slit. The cosmic rays were removed
using the algorithm of \citet{vDo01}, which rejects cosmic rays using a
variation of Laplacian edge detection. The wavelength calibration was
determined from the Ne-Ar lamp spectra.

One-dimensional spectra were extracted using the closest physical
aperture size to the same equivalent circular aperture diameter used
in K02, namely 1.08\,kpc. This aperture will not subtend the same
fraction of all our sample galaxies. Unfortunately not all our sample
galaxies have measured effective radii R$_e$ so that apertures with
the same relative size could not be extracted. This drawback will not
affect our conclusions, however. Using the SAURON integral-field
spectrograph, \citet{Cap06} and \citet{Kun05} have accurately
estimated the effect of aperture corrections on the central velocity
dispersion, the H$\beta$, Mg\,$b$, and Fe5015 indices, after
correcting them for gas emission. If we assume that the Fe5270 and
Fe5335 indices are subject to similar aperture corrections as the
Fe5015 index, even considering a quite conservative range of
0.1\,R$_e$--0.6\,R$_e$ for the relative size of our 1.08\,kpc
aperture, the central velocity dispersion and indices we
measured should not be subject to fluctuation larger than 10\%. This
value is typically smaller or at most comparable to our error bars.

Finally the relative continuum shape of our spectra was
flux-calibrated using the spectrophotometric standard star EG131
following \citet{Bacon01} and \citet{Kun05}, and multiple exposures
of the same galaxies were co-added.

\subsection{Central velocity dispersions and emission-line removal}

In order to correct the line-strength indices for kinematical
broadening and to account for emission-line contamination, we first
measured the recession velocity $V$, and central velocity dispersions
$\sigma$, of our sample galaxies using the method of \citet{Cap04}, and
then applied the procedure described in \citet{Sar05} to derive and
remove the emission-line fluxes if emission was detected.
The algorithm of \citet{Cap04} minimizes the impact of
template-mismatch on the derived stellar kinematics by combining a
number of stellar templates to optimally match the galaxy spectra.
We used templates from the new library of SSP models from
\citet{Vaz05}, which are based on a collection of around a thousand
stellar spectra. The models range from 3500 to 7500\,\AA\ with a
resolution of $\sim$2.3\,\AA. We used SSP models with ages between 1.00
to 17.78\,Gyr and with metallicities of 0.2,
0.,-0.38,-0.68,-1.28,-1.68.
We extracted the stellar kinematics in the wavelength range from 4700
-- 5600\,\AA\ , while masking the regions potentially affected by \Hb\ 
and \Oiii\ emission. Emission from the \Ni$\lambda\lambda$5198, 5200
doublet was never detected.

\placefigtwo 
\placefigthree
Once the stellar kinematics is constrained we removed the mask in the
\Hb\ and \Oiii\ regions and, following the approach of \citet{Sar05},
fit simultaneously the stellar spectrum and the \Hb\ and \Oiii\
emission lines. This is done by adding a set of Gaussian templates
representing the emission lines to the SSP models library, and by
solving only for the position, width, and amplitudes of the
emission-line templates and the contribution of each SSP template. The
latter are convolved by the stellar LOSVD previously derived.
To minimize the impact of template mismatch in the \Hb\ region, we
imposed on the \Hb\ lines the same kinematics as the \Oiii\ doublet,
which was derived in a first emission-line fit.
We then measured the standard deviation in the residuals of our fit
and deemed detected only emission-lines with Gaussian amplitudes at
least 4 times larger than the noise.
Out of the 22 galaxies in our sample, \Hb\ and/or \Oiii\ emission was
detected in 7 galaxies. We subtracted the detected \Hb\ and \Oiii\ 
lines from the spectra, to produce ``emission-line free'' data for the
line-strength analysis.

We note that the relative strength of the \Hb\ to \Oiii\ lines spanned
a wide range of values, from zero to \Hb/\Oiii\ $\ge$ 2.  This suggests
that when possible it is better not to rely on fixed \Hb/\Oiii\ ratios,
such as the \Hb/\Oiii\ $=$ 0.6 values proposed by \citet{Tra98}, to
correct the \Hb\ absorption index.

We applied the same velocity dispersion and emission-line correction
also to the spectra of K02, in order to have all low-density sample
galaxies reduced in the same way.  

\subsection{Line-strength indices}

We measured the Lick/IDS indices as defined in \citep{Wor97,Tra98}. As
our data have higher spectral resolution than the spectra used by the
Lick group, we needed to degrade our spectra in order to measure the
indices, recognizing that we may loose information. We broadened the
spectra with a Gaussian of wavelength-dependent width as the image
dissector scanner (IDS) resolution depended on wavelength. We then
corrected the indices for velocity broadening using the method of
\citet{Kun00} using templates stars and the values for $\sigma$
derived in the previous section. These two steps are needed before
comparing our indices with the stellar population models. We did
not apply any offset to our indices, since in general the typical
offsets are relatively small \citep[e.g.,][]{Nor06} and the index
values measured for the two galaxies from the Lick sample that we
observed were very similar to the published values. We measured the
age-sensitive Balmer indices H$\beta$, and metallicity sensitive
indices Mg\,$b$, Fe5270, Fe5335 among others. The errors
associated to the line-strength indices are obtained by Monte-Carlo
simulations, accounting for the statistical fluctuations in the
spectra and for the uncertainties in the derived recession velocity
and therefore on the position of the line and pseudo-continuum
passbands. The indices and their errors, including CN$1$, CN$2$ and
Ca4227, are given in Table~\ref{index-1}.
Recently \citet{Kun04} found that H$\beta$ is also sensitive to both
the h$_3$ and h$_4$ higher moments of the line-of-sight velocity
distribution (LOSVD), while the Fe5070, Fe5335, Mg\,$b$\/ indices are
sensitive to h$4$ only. For old stellar populations and
$\sigma=250$\,\kms, h$4$ values of $\pm$0.1 can introduce variations as
high as 15 -- 20 \% in the ages and metallicities of old stellar
populations, respectively.

\placefigfour

We also measured the h$_3$ and h$_4$ moments of the central LOSVD and
found that in none of the sample galaxies the values for h$_3$ and
h$_4$ were significant enough to modify the indices (in average
h$_3=-0.02\pm0.03$ and h$_4=0.02\pm0.04$).

\section{Results}

In this section we first analyze index {\em versus} central
velocity dispersion measurements for the $<$Fe$>$, Mg\,$b$\/ and
H$\beta$ Lick/IDS-indices. Then, using the TMB03 models, we
investigate the luminosity-weighted mean age, metallicity and
[$\alpha$/Fe] ratios of the stellar populations. Finally, we explore
the relations between these three parameters and the galaxy mass, as
traced by $\sigma$.

\subsection{Lick/IDS indices as function of velocity dispersion}

\citet{Bur88}, \citet{Jor97} and \citet{Tra98} showed that the
Mg$_2$ and H$\beta$ indices vary systematically with the velocity
dispersion of galaxies $\sigma$, whereas the $<$Fe$>$ index presents a
rather weak correlation with $\sigma$ \citep[see also][]{Kun01}.  We
explored the correlation between line strength and the velocity
dispersion, considering log $<$Fe$>$ {\em versus\/} $\sigma$,
Mg\,$b\,^\prime$ {\em versus\/} $\sigma$ and H$\beta^\prime$ {\em
versus\/} $\sigma$. The prime sign $[^\prime]$, is the index expressed
in magnitudes like the ``molecular'' indices \citep[see
e.g.,][]{Kun01}.

Figure~\ref{indices_sig_cluster} illustrates these relations for
early-type galaxies from the high-density environment sample from
\citet[][hereafter T05]{Tho05}. This sample includes 11 galaxies from
the Virgo cluster \citep{Gon93}, 32 from the Coma cluster
\citep{Meh00,Meh03}, and 11 from \citet{Beu02}. The latter objects were
selected from the ESO-UV Catalog \citep{Lau89} requiring a local galaxy
surface density NG$_T$ $>$ 9 (NG$_T$ being the number of galaxies per
square degree inside a radius of 1 degree around the considered
galaxy).  

There is a good agreement between the relations shown in
Figure~\ref{indices_sig_cluster} and the literature
\citep[e.g.,][]{Coll99,Kun01,Meh03} for both the tight Mg-$\sigma$
relation, and the anti-correlation between H$\beta$ and $\sigma$. Few
galaxies have velocity dispersions below $\sigma=150$\,\kms. There is
no clear distinction between the S0's and E's, except for the
H$\beta^\prime$ vs $\sigma$ relation, where at a given $\sigma$ the
S0's are offset from the E's by $\approx 0.01$\,dex towards higher
values.

Figure~\ref{indices_sig_ours} shows the same diagrams for our sample
on the left and for the low-density environments of K02 (hereafter,
LDE) on the right. Besides the $<$Fe$>$--$\sigma$ relation of isolated
galaxies being steeper than for the other two samples, 
%and the
%H$\beta^\prime$-$\sigma$ of the LDE sample being shallower, 
the relations for the cluster, isolated and LDE samples have similar
slopes within the uncertainties. The scatter in the relations derived
for the isolated sample is however almost twice larger than in the
case of the cluster or LDE samples. 
In detail, the rms for the $\log <$Fe$>$--$\log \sigma$ relation
is 0.050 for the 2dFGRS/Colbert's sample, whereas it is only 0.023 and
0.035 for the cluster and LDE sample, respectively. Similarly, for the
Mg\,$b\,^\prime$--$\log \sigma$ relation it is 0.017 for the
2dFGRS/Colbert's sample, compared to a rms of 0.010 for the cluster
environment and a rms of 0.009 for the LDE sample. Finally, for the
H$\beta\,^\prime$--$\log \sigma$ relation we find a rms of 0.012,
0.009 and 0.008 for the 2dFGRS/Colbert's, the cluster and the LDE
samples, respectively.

We note the distinct grouping in the velocity dispersion distribution
of our sample with one group clustered around 150 km s$^{-1}$ and
another around 250 km s$^{-1}$. This separation may be an artifact of
our sample selection. No such dichotomy is present in our absolute
magnitude distribution. Relatively more galaxies in our sample have a
low velocity dispersion (log $\sigma \, \le \, 2.2$) than the cluster
galaxies.

\placefigfive

In summary, the relations between line strength index and central
velocity dispersion are on average quite similar for cluster galaxies
and both samples of galaxies in low-density environments, following
the same trends: negative correlation between H$\beta$ and the
velocity dispersion, increasing metal line strength (Mg\,$b$,
$<$Fe$>$) with increasing $\sigma$. However, we find a larger scatter
for the early-type galaxies in low-density environments.  Similarly,
\citet{Den05a} found no significant differences in the index--$\sigma$
relations between cluster and low-density region galaxies, except for
a larger scatter for the group, field, and isolated galaxies compared
to the cluster galaxies.

\subsection{H$\beta$ versus [MgFe]$^\prime$}
In this section we aim to derive robust and first order estimates
of the luminosity weighted age and metallicity of our sample
galaxies. A detailed investigation, including the abundance ratios, is
presented in Section~\ref{sec:proctor}. The luminosity-weighed age of
a population, can be inferred from a comparison of selected
line-strength indices with models of single stellar populations such
as those of TMB03. In our study we use H$\beta$ as age indicator
because its age sensitivity is greater and it is less degenerate with
metallicity, or abundance ratio variations, than the H$\gamma$ index
\citep[see e.g.,][]{Korn05}. As a metallicity indicator we adopt the
[MgFe]$^\prime$ index of TMB03 as it is virtually independent of
abundance ratio variations.

Figure~\ref{MgFep_Hb_cluster} shows the distribution in H$\beta$ versus
[MgFe]$^\prime$ for the cluster galaxies and a grid from the stellar
population models of TMB03. The almost vertical lines are of constant
metallicity and range from -2.25 to 0.67 dex, whereas the solid lines
represent the age from 1 to 15\,Gyr.
We note that cluster galaxies appear to have a quite small range
in metallicity for galaxies older than 5\,Gyr; they are almost all
between [Z/H] $=$ 0 and 0.35 dex. Most of these galaxies are
ellipticals. On the other hand the younger galaxies, mainly S0's, have
a larger spread in metallicity reaching [Z/H] $=$ 0.67 dex.

Figure~\ref{MgFep_Hb_ours} presents the same age-metallicity
diagnostic plot but for the samples of field galaxies: on the left
panel, our sample of isolated galaxies and on the right panel, the LDE
sample of K02. Our isolated sample shows a wider spread in ages and
metallicities than in the case of cluster galaxies. Most of our sample
galaxies have ages between 3 and 15\,Gyr and metallicities between
-0.3 and 0.3. The LDE sample of K02 shows also a similar range of ages
and metallicities. Our data is consistent with the low-density
region galaxies of \citet[][hereafter D05b]{Den05b}, which also show a
large spread in age and metallicity. Taking the isolated and LDE
samples together and considering the uncertainties in the Hubble
classification of our sample, we do not find a clear relation
between galaxy type and age or metallicity.
On the other hand, early-type galaxies in clusters tend to have
different properties depending on their type, S0's rather younger with
a large range in metallicity and elliptical galaxies rather older with
a small range in metallicity.

\subsection{$<$Fe$>$ versus Mg\,$b$}

 In this section we have a first look at the abundance ratios of
our sample galaxies. We investigate the Mg\,$b$ {\em versus \/}
$<$Fe$>$ diagram (Figures~\ref{Mg_Fe_cluster}, \ref{Mg_Fe_ours}),
where the model prediction for solar abundance ratios spans a narrow
region in parameter space. This diagram is a diagnostic for the star
formation time scale. Magnesium (like the other $\alpha$-elements) is
formed in the explosion of Type II supernovae which occur rapidly
after a burst of star formation whereas the iron peak elements
originate in Type Ia supernovae which lag behind by at least 1\,Gyr
\citep{Nom84,Woo95}.  Thus a super-solar $\alpha$-element to Fe ratio
indicates that stars formed in an initial burst taking up the Type II
supernova abundance pattern and then star formation ceased, perhaps
through the onset of galactic winds sweeping away the gas material
\citep{Gre83,Mat86,Tho98}.  These short star formation time-scales
$\le$\,1\,Gyr are also interesting as they are not achieved by current
models of hierarchical galaxy formation \citep[e.g.,][]{Mat94, Tho99}.

In Figure~\ref{Mg_Fe_cluster} we plot galaxies in clusters. Most
appear to have [$\alpha$/Fe] ratios between 0 and 0.3 dex, and thus
have on average clearly super solar abundance ratios, a result
first found by \citet{Wor92}. We note that the lenticular galaxies
appear, in this diagram, to be closer to solar abundance ratios than
the elliptical galaxies. However, this could be partly attributed to
an age effect since the models are not completely degenerate in this
parameter space (see Section~\ref{sec:proctor} for a more rigorous
analysis).

Figure~\ref{Mg_Fe_ours} shows the same distribution but for our sample
(left panel), and the LDE galaxies (right panel). The plots are
very similar, there is no obvious difference between ellipticals and
S0s (in contrast to the clusters galaxies) or between our
2dFGRS/Colbert sample of isolated galaxies and the LDE sample of K02.
Most isolated and LDE galaxies display a clear $\alpha$-enhancement,
although at lower level than the cluster galaxies. 
 
\placefigsix

The Mg\,$b$\/ {\em versus}\/ $<$Fe$>$ diagram can be used to obtain a
first estimate of the abundance ratios but the result crucially
depends on the age and metallicity assumed. Often these quantities are
estimated using diagrams opposing two indices that are intended to be
sensitive only to age and metallicity, such as the H$\beta$ versus
[MgFe]$^\prime$ diagrams of \S4.2. However, a perhaps more
economic and precise way of deriving age, metallicity and abundance
ratio estimates is the $\chi^2$ fitting approach pioneered by
\citet[][hereafter P04]{Pro04}. We will explore this technique in the
next section.
  
\subsection{Determination of age, metallicity and abundance ratios}
\label{sec:proctor}

P04 pointed out that diagrams using only a few indices such as
H$\beta$ or H$\gamma$ {\em versus}\/ [MgFe] \citep[see e.g.,][]{Gon93}
can lead to inconsistent age estimates, even when used to derive the
age of globular clusters.  In order to improve the stellar population
analysis P04 proposed to solve ($\chi^2$ minimization) for the best
fitting stellar population parameters using a large set of
indices. This new method is making better use of the information on
the age, metallicity and abundance ratio which is contained in all the
indices rather than being biased by a few selected indices.

In our application of the method we interpolated all tabulated indices
from TMB03 in steps of 0.025 in metallicity, 0.05 in [$\alpha$/Fe] and
1\,Gyr in age. The actual $\chi^2$ minimization was performed in the
log age, [Z/H] and [$\alpha$/Fe] parameter space.  Among the indices we
measured, we decided to use only G4300, Fe4383, H$\beta$, Fe5015,
Mg\,$b$, Fe5270, and Fe5335 in the analysis. We excluded the CN1, CN2
and Ca4227 indices because they are not well reproduced within the
TMB03 models \citep[][P04]{Tri95,Tho03b}.  Errors on the final
estimates were obtained by Monte-Carlo simulations using the errors on
the index measurements.

\placefigseven

The right panel of Figure~\ref{proctor_plots_cluster_ours} shows our
final results for the isolated galaxies in our sample together with the
galaxies in the LDE sample of K02, since these samples have very
similar properties.  The left panel of
Figure~\ref{proctor_plots_cluster_ours} shows the same plots for the
cluster galaxies of \citet{Tho05}. A similar figure was presented by
T05, but included only galaxies older than 5\,Gyr, i.e. their old
sub-sample. T05 obtained their values by first fixing the abundance
ratio and determining the ages and metallicities with the two pairs
(H$\beta$, Mg\,$b$), (H$\beta$, $<$Fe$>$), then used the metallicities
found for Mg\,$b$\/ and $<$Fe$>$ to adjust the abundance ratio, and
iterated until consistent results between both pairs were found.
Table~\ref{index-1} gives our age, metallicity and abundance
ratios estimates and associated errors for the low-density environment
galaxy sample. The full table is available in the electronic version
of MNRAS.

\placefigeight
\placetabthree

If we look first at galaxies in clusters (left panel), we notice the
clear correlation between [$\alpha$/Fe] and $\log \sigma$ found also
by e.g., K00, K01, and T05. The three top plots show that old galaxies
have also higher velocity dispersions and higher abundance ratios than
younger objects. Similarly, \citet{Cal03} and \citet{Tra00b} claim
that there is a trend for lower $\sigma$ elliptical galaxies to have
younger ages. Using a Spearman rank order test between $\sigma$ and
age, we find that ellipticals follow a better trend of $\sigma$ with
age than S0s (98\% probability for the existence of a
correlation for Es and 85\% probability for S0s) as on average the
scatter for the S0s is larger. Furthermore, on average Es are older
than S0s, with mean luminosity-weighted ages of 11 and 6\,Gyr for Es
and S0s, respectively.  From the plots against the metallicity, the
cluster galaxies appear to have only super-solar metallicities, with
an average [Z/H] $\approx0.27$, and no  clear trend with $\sigma$
or abundance ratio. There also appears to be a weak
anti-correlation between [Z/H] and age.
Spearman rank order tests yield a 60\% and 95\% probability that a
correlation exists for Es and S0s, respectively.  However, the
true significance of such a relation is difficult to assess, since the
errors in age and metallicity are correlated in the direction of the
potential relation \citep[see e.g,][for details]{Kun01,Tho05}.
Correlations between $\sigma$ and age, [Z/H], and [$\alpha$/Fe] for
cluster galaxies were also found by T05 (but see also references
therein).

In contrast, for the isolated and LDE galaxies (right panel), we
notice in all the plots a large spread in age, metallicity and
abundance ratio.
A Spearman rank order test suggests that metallicity is
correlated with $\sigma$ or age with a probability of 95\%, although
as in the case of the cluster galaxies, the relation between metallicity
and age would need to be confirmed.

%A weak relation between age and metallicity is found also in cluster
%galaxies, suggesting that younger galaxies tend to have higher
%metallicity than older galaxies irrespective of their environment. The
%same trend was observed in galaxies in different environments by D05b.

We note that contrary to this study both K02 and T05 found evidence for
slightly higher [Z/H] in field early-type galaxies than in clusters.
The discrepancy with the results of T05 may be due to a different
sample selection. Most of the field galaxies of T05 are not as
``isolated'' as the ones in our and K02 samples. On the other hand, the
difference between the conclusions on the stellar metallicity reached
by us and K02 arise from a) our use of a more accurate correction for
the emission-line contamination, b) re-analyzing the data of K02 with
the latest SSP models, and c) from the fact that the cluster of
reference in K02, Fornax, have different population properties than the
Virgo and Coma clusters from T05 used here.

The large scatter we find for the luminosity weighted ages of
early-type galaxies in low density environments is in agreement with
the predictions of semi-analytical models \citep[e.g., ][see also
K02]{Cole00}. Additionally, some of the early-type galaxies with
velocity dispersions of about 150\,\kms\/ reach solar abundance ratios
which would be expected in a hierarchical formation scenario with
extended star formation episodes for galaxies in low density regions
\citep[e.g.,][]{Tho99b}. However, we still find that the most massive
galaxies (i.e. highest velocity dispersion) in the low density
environments show similar abundance ratios as cluster galaxies and
thus have super solar ratios which is difficult to explain in
hierarchical formation scenarios. Similarly puzzling is also the
finding of super solar abundance ratios in less massive galaxies,
which was already noticed by D05b.

Finally, we note that ellipticals and S0s appear to be
indistinguishable in low-density environments, which could reflect our
inability to separate elliptical from lenticular galaxies on the basis
of DSS images. On the other hand, only one third of the combined
isolated and LDE samples, were classified in this way and the majority
of the objects have well defined morphologies. Hence elliptical
galaxies and S0s in low-density environments appear to have similar
stellar populations, in contrast to what is observed in clusters.
  
Overall, galaxies in low-density environments exhibit a more
uniform distribution of luminosity-weighted stellar age and a broader
range in metallicity than their counterparts in clusters, and on
average also appear to be younger. Galaxies in low-density
environments can achieve the same degree of non-solar abundance ratios
as galaxies in clusters, although a larger range of [$\alpha$/Fe]
ratios, even reaching solar values at intermediate velocity
dispersions, is observed. The large range in stellar population
parameters and lack of correlation with central velocity dispersion
observed in our sample is indicative of a more diverse galaxy formation
scenario in the field as compared to the typical cluster environment.

\section{Conclusions}
Following the pilot project of \citet{Kun02} we set out to study the
stellar populations of a larger and more accurately selected sample of
galaxies in low-density environments. We have used spectroscopic data
from the 2dFGRS survey to select galaxies without strong H$\alpha$
emission and with no more than one and five companion within 1 and
2\,Mpc$^{-1}$, respectively, and combined these objects with isolated
early-type galaxies from the sample of \citet{Colb01}, which have no
companions within 1\,Mpc.

We measured a set of Lick/IDS indices to obtain estimates of the
  luminosity-weighted mean age, metallicity, and abundance ratio in our
  sample galaxies. We also applied a new technique \citep{Sar05} to
  subtract gaseous emission from our spectra before
  measuring the absorption line-strength indices. The stellar
  population estimates were obtained with the models of \citet{Tho03a}
  and the $\chi^2$ technique of \citet{Pro04}.
 
By applying the same methodology to the data of K02, we find that the
stellar population properties of the galaxies of K02 appear to be very
similar to that of our sample galaxies. After combining all
low-density environment samples together (yielding a sample of 31
galaxies), we find that, compared to the cluster galaxies from the
compilation of \citet{Tho05}, early-type galaxies in low-density
environments:

\begin{enumerate}
  
\item Show a more uniform distribution of luminosity weighted stellar
    ages ranging from $\sim$2 to 15\,Gyr and thus have on average
    younger stellar populations. This is particularly true for
    elliptical galaxies.
  
\item Display a broader range of stellar metallicities, extending to
  sub-solar values at similar central velocity dispersions, $\sigma$.
  
\item Can also reach the same amount of non-solar abundance ratios as
    cluster galaxies. However, there is an increased scatter at fixed
    $\sigma$ with no clear correlation between [$\alpha$/Fe] and
    $\sigma$.
  
\item Elliptical galaxies and S0s in low-density environments have
    stellar populations that are indistinguishable within our sample
    in contrast to the cluster galaxies, where there is a significant
    distinction between younger S0s and older elliptical galaxies.

\end{enumerate}

How do these results compare with the predictions of galaxy-formation
models \citep[e.g.,][]{Cole00}? According to the models, galaxies in
clusters assemble early, when they repeatedly experienced short
star-formation episodes each time they merged. More recently, however,
galaxies in clusters do not merge owing to their high relative
velocities. On the other hand, galaxies in the field initially merge
less often \citep[e.g.][]{Kho01}, while merging continues to low
redshifts.

Hence, even considering only merging-induced star formation, early-type
galaxies in low-density environments are expected to show a wider range
of ages for their stellar constituents, compared to the stars in
cluster galaxies. This is qualitatively consistent with our findings as
also previously reported by K02 and T05.

In addition, the larger range of [$\alpha$/Fe] and metallicities
observed in field early-type galaxies suggest further differences
between the evolutionary paths of cluster and field galaxies.
The large range of [$\alpha$/Fe] ratios reaching solar values
indicates that at least some field galaxies experienced extended star
formation episodes. We can speculate that these stars formed in the
disks that grew between merger events \citep{Kho05}.
Finally, we note that the ability of galaxies in low-density
environments to grow stellar disks around them in recent epochs may
explain why early-type galaxies in low-density environments are so
scarce and very often show morphological signatures \citep[such as
outer shells or dusts,][K02]{Colb01} of recent merging events. In other
words, most galaxies in the field that we classify as E or S0 may have
assembled only recently.

\section*{Acknowledgments}
We wish to thank Sadegh Khochfar and Daniel Thomas for many helpful
discussions.

\label{lastpage}
\end{document}